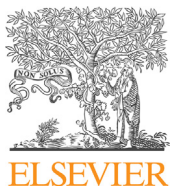
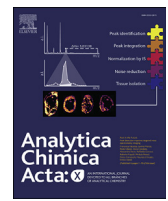

Contents lists available at ScienceDirect

# Analytica Chimica Acta: X

journal homepage: www.journals.elsevier.com/analytica-chimica-acta-x

# Extraction of hydrophobic analytes from organic solution into a titanate 2D-nanosheet host: Electroanalytical perspectives

Wulan Tri Wahyuni [a, b], Budi Riza Putra [a, b], Christian Harito [c], Dmitry V. Bavykin [c], Frank C. Walsh [c], Philip J. Fletcher [d], Frank Marken [a, *]

[a] Department of Chemistry, University of Bath, Claverton Down, BA2 7AY, Bath, UK
[b] Department of Chemistry, Faculty of Mathematics and Natural Sciences, Bogor Agricultural University, Bogor, West Java, Indonesia
[c] Energy Technology Research Group, Faculty of Engineering and Physical Sciences of Southampton, SO17 1BJ, Southampton, UK
[d] Materials and Chemical Characterisation Facility (MC2), University of Bath, Claverton Down, BA2 7AY, Bath, UK

A B S T R A C T

Titanate nanosheets (single layer, typically 200 nm lateral size) deposited from aqueous colloidal solution onto electrode surfaces form lamellar hosts that bind redox active molecular redox probes. Here, hydrophobic redox systems such as anthraquinone, 1-amino-anthraquinone, deca-methylferrocene, 5,10,15,20-tetraphenyl-21$H$,23$H$-porphine manganese (III) chloride (TPPMnCl), and α-tocopherol are shown to bind directly from cyclopentanone solution (and from other types of organic solvents) into the titanate nanosheet film. For anthraquinone derivatives, stable voltammetric responses are observed in aqueous media consistent with 2-electron 2-proton reduction, however, independent of the pH of the outside solution phase environments. For decamethylferrocene a gradual decay of the voltammetric response is observed, but for TPPMnCl a more stable voltammetric signal is seen when immersed in chloride containing (NaCl) electrolyte. α-Tocopherol exhibits chemically irreversible oxidation and is detected with 1 mM–20 mM linear range and approximately $10^{-3}$ M concentration limit of detection. All redox processes exhibit an increase in current with increasing titanate film thickness and with increasing external electrolyte concentration. This and other observations suggest that important factors are analyte concentration and mobility within the titanate host, as well as ion exchange between titanate nanosheets and the outside electrolyte phase to maintain electroneutrality during voltammetric experiments. The lamellar titanate (with embedded tetrabutyl-ammonium cations) behaves like a hydrophobic host (for hydrophobic redox systems) similar to hydrophobic organic microphase systems. Potential for analytical applications is discussed.

## 1. Introduction

Modified electrodes provide an important tool in sensor development [1]. There is considerable interest in new types of 2D-nanosheet materials with new properties that can contribute to the development of new analytical procedures based on modified electrodes. Recent progress has been made, for example, in the development of graphene derivatives [2], MXenes [3], metal chalcogenides [4,5], and other types of lamellar 2D-oxides [6]. Nanosheet architectures provide a high level of surface exposure and inter-lamellar spaces, within which catalysts can be immobilised [7], catalytic reactions can occur [8], or analytical detection can be performed [9]. Tong and coworkers demonstrated catalytic nitrite sensing with an iron porphyrin/titanoniobate nanosheet composite modified electrode [10]. In a recent study, we investigated the immobilisation and reactivity of ferrocene boronic acid as molecular receptor for glucose or fructose in titanate nanosheet hosts [11]. Titanates and titanate nanosheets provide an important class of materials [12,13] and they are highly negatively charged. Binding and exchange of cations is a well known pattern of reactivity for titanates [14]. Films can be deposited onto electrodes directly from aqueous colloidal solution. The inorganic nature of the titanates may suggest that inter-lamellar space in titanate nanosheet deposits should be highly polar and hydrophilic. However, it is shown here that titanate deposits in fact show behaviour more closely related to hydrophobic hosts. It is suggested that the nature of the associated cations (here tetrabutylammonium in the titanate film) could be a factor affecting the reactivity and swelling of titanate nanosheet coatings when immersed in different types of liquid media.

* Corresponding author.
  E-mail address: f.marken@bath.ac.uk (F. Marken).

Titanate nanosheet materials have been pioneered and developed by Sasaki and coworkers [15–17]. A molten salt reaction was employed to produce an intermediate lepidocrocite-like caesium titanate ($Cs_{0.7}Ti_{1.825}\square_{0.175}O_4$), which after cation exchange in acid and alkylammonium (e.g. tetrabutylammonium) cations exfoliates to smectitite-like acidic titanate of composition $H_{0.7}Ti_{1.825}\square_{0.175}O_4H_2O$) [18]. A colloidal solution is obtained with sheets of typically 200 nm size and a nominal thickness of 1.3 nm or a single unit cell [19]. This colloidal material is readily converted into new hybrid materials [20] and nano-composites [21], and when deposited onto glassy carbon electrodes, it gives films with lamellar structure. Lamellae only merge back into solid bulk structures, when heated to temperatures much higher than ambient. This titanate coating has the ability to bind and retain inter-lamellar redox active molecules and it allows ion uptake and transport in between nanosheets to sustain redox processes at the electrode surface (Fig. 1).

Ion transport and membrane properties of 2D nanosheet materials are of considerable interest [22]. The presence of negative charge on the titanate (as indicated by the negative zeta potential) requires the presence of counter cations (here indicated as tetrabutylammonium cations). This introduces ion conductivity and exchange as well as a degree of semi-permeability depending on the interaction with the external electrolyte. When immobilising a redox active molecule into the titanate layers and when applying a potential to switch the redox state (immersed in aqueous electrolyte), ion transfer or exchange at the titanate nanosheet | aqueous electrolyte interface is required to balance the charges (to maintain electroneutrality). For related microphase redox processes (e.g. based on immobilised oil microdroplets at electrode surfaces [23]) coupled electron transfer and ion exchange has been investigated with voltammetric methods. Conditions within the microphase deposit depend on the type of cation or anion that transfers at the interface. It is shown here that also for titanate nanosheet deposits ion exchange across the titanate|aqueous electrolyte interface is important.

Here, the behaviour of titanate nanosheet deposits is investigated for several hydrophobic redox systems. Anthraquinone, 1-amino-anthraquinone, decamethylferrocene, tetraphenylporphyrinato-Mn(III)chloride (TPPMnCl), and α-tocopherol are extracted directly from an organic solution phase into the titanate film (deposited on a glassy carbon electrode). The type of organic solvent and the concentration are shown to affect the extraction process and the resulting voltammetric behaviour. The effects of aqueous electrolyte and pH are considered and a model is developed to explain the behaviour of these immobilised redox systems. The process of extraction of analyte from organic solution followed by detection in aquoues electrolyte could be of interest in analytical procedures. However, both the fundamental understanding and analytical sensitivity of processes will have to be further improved.

## 2. Experimental

### 2.1. Chemical reagents

Titanate nanosheet material was synthesized as describe previously by Sasaki [24] and by Harito et al. [20]. Anthraquinone (molecular weight 208.21 g mol$^{-1}$; CAS: 84-65-1; 97%), 1-aminoanthraquinone (molecular weight 223.23 g mol$^{-1}$; CAS: 82-45-1; 97%), decamethylferrocene (molecular weight 326.30 g mol$^{-1}$; CAS: 12126-50-0; 97%), 5,10,15,20-tetraphenyl-21$H$,23$H$-porphine manganese (III) chloride (TPPMnCl; molecular weight 703.11 g mol$^{-1}$; CAS: 32195-55-4; 95%), and ±α-tocopherol (molecular weight 430.71 g mol$^{-1}$; CAS: 10191-41-0; ≥96%) were obtained from Sigma-Aldrich. Sodium dihydrogen phosphate and sodium phosphate dibasic hexa-hydrate (for buffer preparation), sodium chloride, cyclopentanone, acetone, chloroform, dichloromethane, and ethanol were purchased from Sigma-Aldrich, Fisher Scientific, or VWR BDH Chemicals and used without further purification. Solutions were prepared under ambient condition in volumetric flasks with ultrapure water with resistivity of 18.2 MOhm cm (at 22 °C, from an ELGA Purelab Classic system).

### 2.2. Instrumentation

A classic three-electrodes system controlled with a micro-Autolab III system (Metrohm-Autolab, Netherlands) with NOVA 2.1.2 software was used for cyclic voltammetry measurement. The counter and reference electrodes were platinum wire and Ag/AgCl in 3 M NaCl (from BASi). The working electrode was a 3.0 mm-diameter glassy carbon electrode (GCE from BASi) modified by coating with a titanate nanosheet film. Characterization of titanate nanosheet materials was carried out using Transmission Electron Microscopy (TEM) analysis (JEOL JEM-2100Plus). The thickness of titanate nanosheet films on glassy carbon was analysed using cross-sectional Scanning Electron Microscopy (SEM, JEOL JSM6480LV) after freeze-fracture. Raman spectroscopy analysis of pure titanate films and film with embedded extracted hydrophobic redox systems were performed on a Renishaw inVia confocal microscope system.

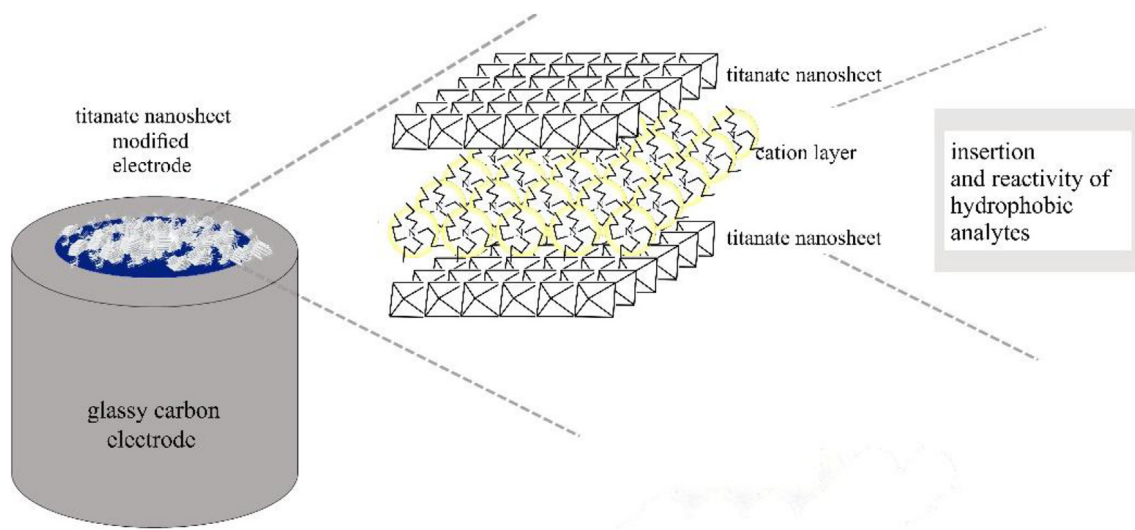

**Fig. 1.** Schematic drawing of the titanate nanosheet exfoliated with tertrabutylammonium cations and re-deposited from colloidal aqueous solution onto a glassy carbon electrode. This deposit is then investigated as sorbent and host for hydrophobic redox systems (bound from organic solution) and for electrochemical reactivity (when immersed in aqueous electrolyte media).

### 2.3. Procedures

**Electrode modification.** The glassy carbon electrode (GCE, 3.0 mm diameter, BSAi) was polished with alumina slurry (Buehler, 0.3 μm) and cleaned with isopropanol and filtered water before modification. An aliquot of 4 or 8 μL titanate nanosheet colloidal solution (2.56 g L$^{-1}$) was deposited onto the electrode surface by drop casting. The solution was dried at ambient temperature forming thin visible (opaque-white) films of titanate nanosheets. Fig. 2 shows TEM data for titanate nanosheets, typical diffraction data for titanate nanosheet assemblies [16], and cross-sectional SEM data for 4 μL and 8 μL deposits with approximately 1.9 μm and 3.3 μm thickness. Electron diffraction pattern are consistent with diffraction data for a lamellar system as reported initially by Sasaki et al. [16]. The distinct widening and stabilisation of the inter-lamellar space with the intercalaction of bulky organic exfoliation cations has been reported [25].

**Extraction of redox-active analytes.** Redox-active analytes solution were prepared in cyclopentanone (or in other types of organic solvents such as ethanol, chloroform, dichloromethane, or acetone). The glassy carbon electrode modified with titanate nanosheet film was immersed into the solutions (for a defined period of time) to allow the extraction of redox-active analytes from its solution into the titanate nanosheet film. After immersion the electrode was dried under ambient conditions.

**Electrochemical and zeta-potential measurements.** Electrochemical measurements were carried out after immersion of the modified electrode into aqueous electrolyte solution. The zeta-potential for the colloidal titanate solution in water was measured on a Zetasizer Nano ZS (Malvern Instruments Ltd, Malvern, UK) for 0.25 gL$^{-1}$ (−33.9 mV)), for 0.025 gL$^{-1}$ (−32.4 mV), and for 0.0025 gL$^{-1}$ (−31.2 mV) to give an average of −32.5 mV consistent with an overall negative charge before deposition.

**Raman measurements.** An aliquot of 4 μL titanate nanosheet colloidal solution (2.56 g L$^{-1}$) was deposited on to silicon wafer surface by drop casting method. The solution was dried at ambient temperature forming thin film of titanate nanosheets. Raman spectroscopy analysis of titanate nanosheet film was performed on a Renishaw inVia system with the following condition, detector: CCD Camera 1040 × 256; Laser: 532 nm; Grating: 1800 mm$^{-1}$ (532 nm); Exposure time: 10 s 30 accumulations; Laser power: 1%; Center: 1033 Raman shift/cm$^{-1}$; Objective ×50.

## 3. Results and discussion

### 3.1. Binding of anthraquinone derivatives into a titanate nanosheet host

Anthraquinone may be regarded as a model hydrophobic redox system with very low solubility in aqueous media and good electrochemical reactivity in contact to aqueous electrolyte [26]. The wider class of anthracenediones is widely used in industry and domestic products and commercially significant [27].

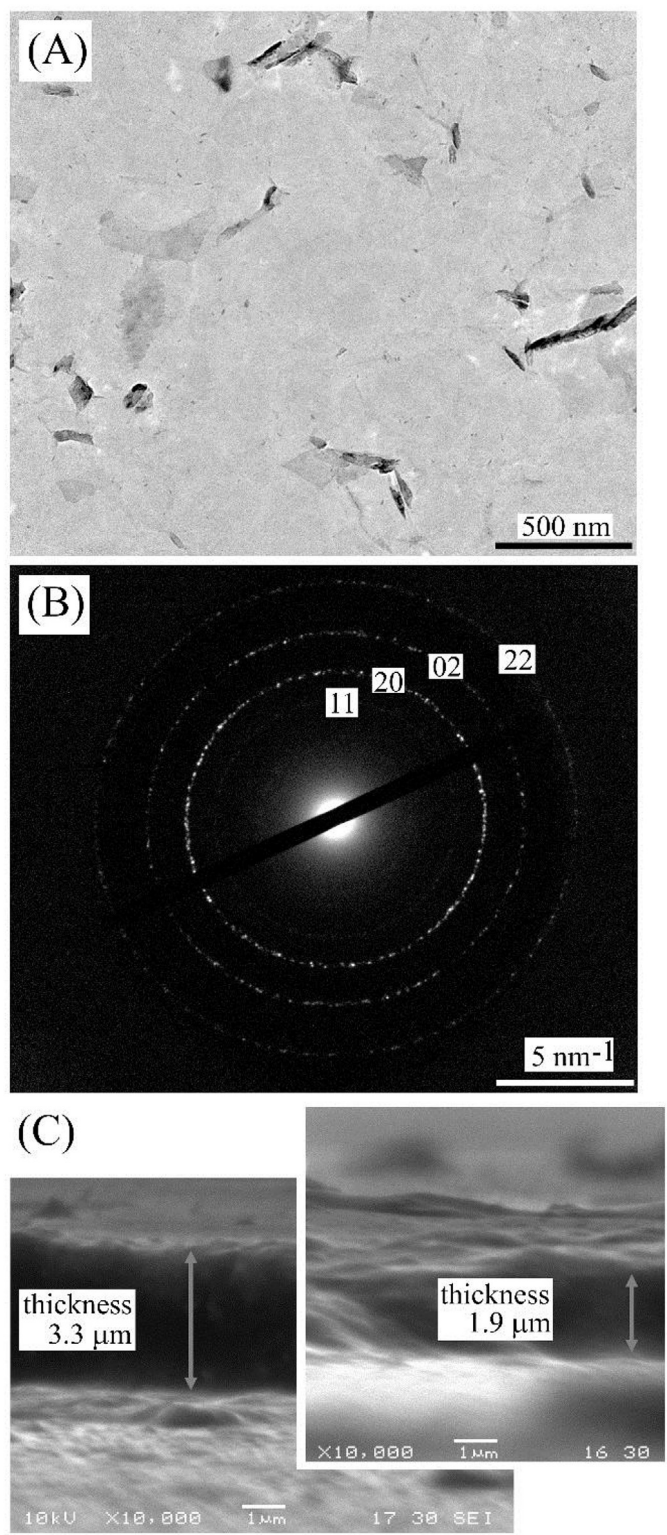

**Fig. 2.** (A) Transmission electron microscopy (TEM) image of titanate nanosheet material. (B) Electron diffraction pattern showing the main diffraction lines for the stacks of nanosheets with approximately 1.8 nm inter-nanosheet gap. (C) Scanning electron microscopy (SEM) images for films deposits of titanate nanosheet films of approximately 1.9 μm and 3.3 μm thickness.

Anthraquinone is reduced in a 2-electron 2-proton process (Equation 1) to the corresponding anthraquinol (see reaction equation in Fig. 3) and has been used, for example, as a probe for local pH in aqueous media [28]. When immobilised at an electrode surface the voltammetric peak feature for the anthraquinone reduction would usually exhibit a Nernstian shift of 2.303 RT/F per pH unit change in aqueous pH (or approximately 58 mV/pH at room temperature or 22 °C [29]).

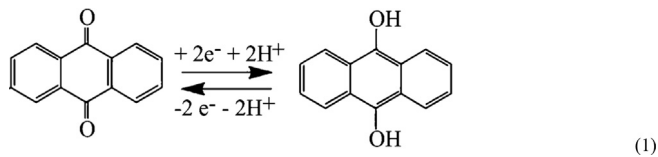

(1)

Fig. 3A shows typical voltammetric responses for anthraquinone immobilised into a film of titanate nanosheets on a glassy carbon electrode. Electrodes were prepared by initially coating titanate nanosheets onto the glassy carbon electrode. This electrode is then immersed for 60 min into a solution of 2.5 mM anthraquinone in cyclopentanone, removed and dried. The resulting modified electrode clearly shows the voltammetric response consistent with the reduction of anthraquinone in aqueous phosphate buffer solution environment. The midpoint potential of ca. −0.43 V vs. Ag/AgCl is consistent with that reported for anthraquinone in aqueous media and also for surface immobilised anthraquinone moities [30,31]. Most reports show a slightly more negative $E_{mid}$ under similar conditions and a Nernstian response to pH changes [32]. However, in the case of titanate immobilised anthraquinone, the pH dependence of the voltammetric signal is suppressed (*vide infra*).

Data in Fig. 3A are shown for two different loadings of titanate nanosheets on the glassy carbon electrode surface (films of approximately 1.9 μm and 3.3 μm thickness) and a clear increase in the peak current for both reduction and oxidation occurs when increasing the titanate film thickness. This was confirmed by additional experiments (not shown) employing deposition of the same mass of titanate nanosheets and extending the deposit over the outside insulator of the electrode (thereby thinning the coating over the electrode). Titanate nanosheet material present outside of the glassy carbon perimeter is inactive and therefore not responsible for voltammetric responses. Therefore, the film thickness for titanate nanosheet deposits on the glassy carbon surface is the key to further enhancement of the voltammetric signal. This observation could suggest that diffusion (transport) of anthraquinone within the film occurs in the given time frame and space. The diffusion layer can be linked to the diffusion coefficient (by dimensional analysis) to give $D_{app} = \frac{vF\delta^2}{RT}$ [33], which can be estimated here (see Fig. 3C) as $D_{app} = 0.04 \times 10^{-9}$ m$^2$s$^{-1}$ or faster. Therefore, the anthraquinone molecule in oxidised and reduced form could have considerable mobility within the titanate nanosheet host material to explain the observations. This considerable level of mobility could be linked to the presence of organic cations within the lamella structure (see Fig. 1). However, more likely in this case is that another species (an electrolyte counter ion) is responsible for the apparent diffusion process and the magnitude of the voltammetric response. Anthraquinone is in fact reactive only very close to the electrode surface with much lower rates of diffusion within the titanate host (*vide infra*).

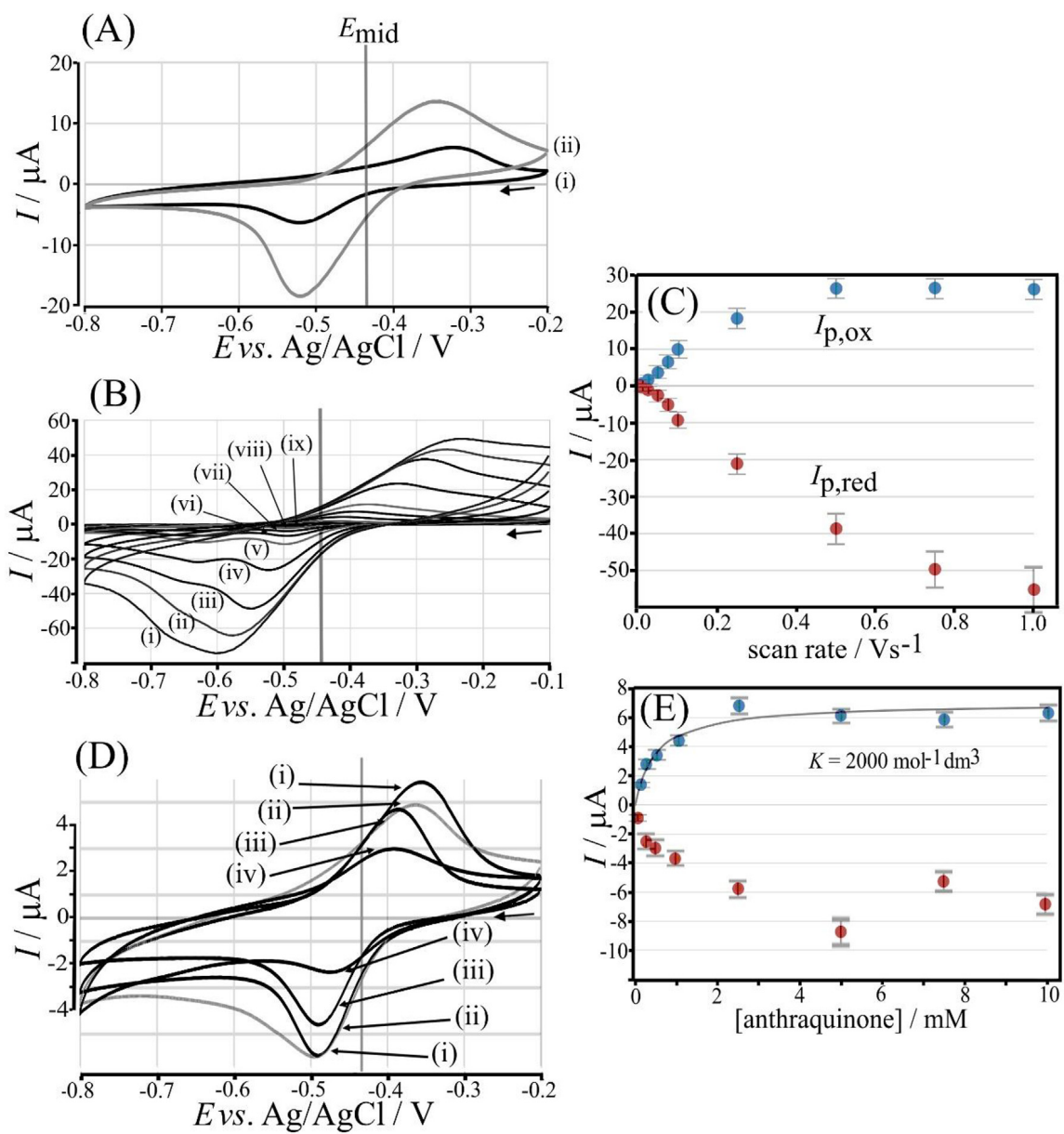

**Fig. 3.** (A) Cyclic voltammogram (scan rate 100 mVs$^{-1}$; in 0.1 M phosphate buffer pH 7) for anthraquinone extracted from 2.5 mM solution in cyclopentanone at (i) 1.9 μm thickness titanate film modified GCE and (ii) 3.3 μm thickness titanate film modified GCE, (B) Cyclic voltammogram (scan rate (i) 1, (ii) 0.75, (iii) 0.5, (iv) 0.25, (v) 0.1, (vi) 0.075, (vii) 0.05, (viii) 0.025, (ix) 0.01 Vs$^{-1}$; in 0.1 M phosphate buffer pH 7) for anthraquinone extracted from 2.5 mM solution in cyclopentanone at 1.9 μm thickness titanate film modified GCE. (C) Plot of anodic and cathodic peak currents *versus* scan rate (estimated error 10%). (D) Cyclic voltammogram (scan rate 100 mVs$^{-1}$; in 0.1 M phosphate buffer (i) pH 7, (ii) pH 9, (iii) pH 10, (iv) pH 11) for anthraquinone extracted from 2.5 mM solution in cyclopentanone at 1.9 μm thickness titanate film modified GCE. (E) Plot of peak current *versus* concentration of anthraquinone in cyclopentanone solution (estimated error 10%). The line shows a Langmuir isotherm fitting with an approximate binding constant of $K = 2000$ mol$^{-1}$ dm$^3$.

When investigating the effect of scan rate (see Fig. 3B and C) in cyclic voltammograms for immobilised anthraquinone, the anticipated increase with square root of scan rate (for a diffusion controlled case) can be seen, but the peak shapes are more complex for this simple relationship to hold. Fig. 3D shows data for the reduction of anthraquinone immobilised in titanate nanosheets at different pH values (all in 0.1 M phosphate buffer and for freshly prepared electrodes). The absence of a significant pH shift is striking and indicative of a mechanism that does not include proton or hydroxide exchange across the titanate nanosheet | aqueous electrolyte phase boundary. Other types of ions may be involved. In this particular case "excess electrolyte anions" within the lamella space are suggested to undergo expulsion during anthraquinone reduction (*vide infra*).

The effect of anthraquinone concentration in the cyclopentanone solution was investigated over a range from 0.1 to 10 mM (see Fig. 3E). The plot of peak current versus concentration reveals an initial increase with concentration followed by a plateau. The data set can be fitted based on the simplistic model of a Langmuir isotherm with a binding constant $K = 2000$ mol$^{-1}$ dm$^3$. However, this interpretation is likely to be wrong and the underlying mechanisms for binding is unlikely to lead to saturation. For both, binding and electrochemical conversion, there is more complexity as is revealed in data shown in Fig. 4. The reason for the plateau in peak currents here is suggested to be linked not to the availability of binding sites, but instead to a limited amount of "excess electrolyte anions" undergoing expulsion into the aqueous phase during anthraquinone reduction. This can explain both, the observed titanate film thickness effects on voltammetric responses and the absence of pH effects on voltammetric signals for anthraquinone.

When investigating the effects of different organic solvents on the anthraquinone binding process (see Fig. 4A), considerable effects are observed when comparing (i) cyclopentanone, (ii) acetone, (iii) chloroform, and (iv) dichloromethane. The process appears to be working particularly well for anthraquinone in cyclopentanone, although reasons for this behaviour are currently not fully understood (although likely to be linked to the presence of tetrabuylammonium cations from the titanate exfoliation process). Swelling of the titanate nanosheets in different types of solvents could be important to allow uptake of redox active species. Even more striking is the effect of the supporting electrolyte concentration in the aqueous phase on the voltammetric response for 2.5 mM anthraquinone absorbed from cyclopentanone (Fig. 4B). With increasing NaCl electrolyte concentration, the voltammetric response systematically increases. This suggests that anthraquinone bound into titanate nanosheet deposit is not fully converted, and that only a thin layer containing anthraquinone close to the surface of the electrode is electrochemically active. The higher NaCl electrolyte concentration may lead to more partitioning of NaCl (and therefore additional chloride anions) into the titanate nanosheet film and this could be the underlying reason for the increase in the current signal (explaining both, the electrolyte concentration effect and film thickness effect). Therefore, a limiting factor in the voltammetric response could be "internal excess electrolyte" within the titanate nanosheet material. The plateau region in the plot in Fig. 3E could be associated with amount of internal free electrolyte (in this case mobile chloride anions that are expelled from the titanate host during reduction). In addition, the observed $D_{app}$ for anthraquinone (*vide supra*) could then be linked to electrolyte mobility and to a lesser extent to mobility of anthraquinone.

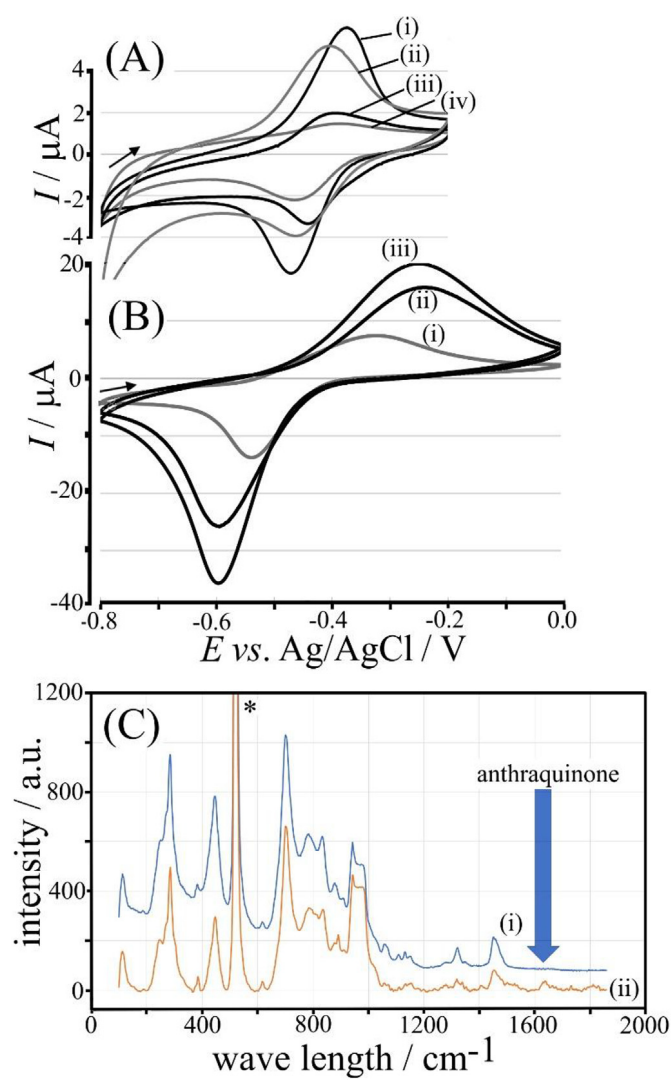

**Fig. 4.** A) Cyclic voltammogram (scan rate 100 mV s$^{-1}$; in 0.1 M phosphate buffer pH 7) for anthraquinone extracted from 2.5 mM solution in (i) cyclopentanone, (ii) acetone, (iii) chloroform, (iv) dichloromethane solution at 1.9 μm thickness titanate film modified GCE, (B) Cyclic voltammogram (scan rate 100 mV s$^{-1}$; in (i) 0.1, (ii) 0.5, (iii) 1 M NaCl) for anthraquinone extracted from 2.5 mM solution in cyclopentanone at 1.9 μm thickness titanate film modified GCE. (C) Raman spectra of pure titanate film (blue) and film with absorbed anthraquinone (orange). The * symbols indicates a silicon substrate signal.

In order to obtain additional evidence for the presence of anthraquinone in the titanate nanosheet film material, additional Raman spectroscopy experiments were performed. Fig. 4C shows data for the pure titanate material and for the film with anthraquinone absorbed from 2.5 mM anthraquinone in cyclopentanone (60 min immersion). It can be seen that several strong Raman peaks are seen at 140, 280, 450, and 705 cm$^{-1}$, consistent with previously reported Raman signals for example for $Cs_xTi_{2-x/4}\square_{x/4}O_4$ [34,35]. Only one new peak associated with the presence of anthraquinone at approximately 1650 cm$^{-1}$ is detected in this wave number range consistent with this being the first stronger Raman signal associated with C=O stretching for anthraquinone [36,37].

Further experimental data were obtained with a coloured derivative of anthraquinone. The anthraquinone derivative 1-amino-anthraquinone is known to be toxic and present in industrial wastes, where it can be detected, for example, by voltammetry or by polarography [38]. The pK$_A$ for the amino group is 1.19 [39] and therefore this molecule remains uncharged and similar in solubility and behaviour compared to that of the parent anthraquinone.

Data in Fig. 5A shows typical voltammograms for the reduction of 1-amino-anthraquinone absorbed from 2.5 mM solution in cyclopentanone in two thicknesses of titanate nanosheet deposits. Very similar behaviour compared to that observed for anthraquinone (see Fig. 3) suggests very similar underlying reactions. The reduction is likely to follow a 2-electron 2-proton pathway (see equation 2). Fig. 5B shows data for voltammetry experiments performed in 0.1 M, 0.5 M, and in 1.0 M aqueous NaCl solution and the increase of current with ionic strength is again clearly observed.

Fig. 5D shows the plot of peak current data *versus* 1-amino-anthraquinone concentration clearly revealing the initial increase followed by a plateau. It seems likely that here again the reduction of 1-amino-anthraquinone is limited by "internal excess electrolyte" rather than by the amount of 1-amino-anthraquinone bound into the titanate nanosheet host material. This is confirmed by the coloration of the titanate nanosheet material after immersion into the red-coloured 1-amino-anthraquinone (see Fig. 5C). With 30 min deposition time a clear coloration develops indicative of a substantial amount of 1-amino-anthraquinone in the titanate nanosheet host. It seems likely that only a small fraction of this material (close to the electrode) is converted in the electrochemical reduction and the voltammetric reduction peaks are associated with chloride expulsion from the titanate nanosheet deposit. Raman data were not obtained due to strong fluorescence of the 1-amino-anthraquinone.

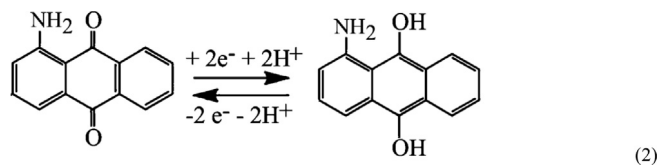

(2)

### 3.2. Binding of decamethylferrocene into a titanate nanosheet host

Next, a wider range of redox active guest species are considered with the common feature of being soluble in organic solution and insoluble in aqueous electrolyte media. Decamethylferrocene offers a 1-electron redox system (see equation 3) with minimal complications from chemical reactivity. Previous studies have shown that decamethylferrocene is water-insoluble and chemically stable and readily oxidised with a (anion-dependent) midpoint potential close to 0.0 V vs. Ag/AgCl [40].

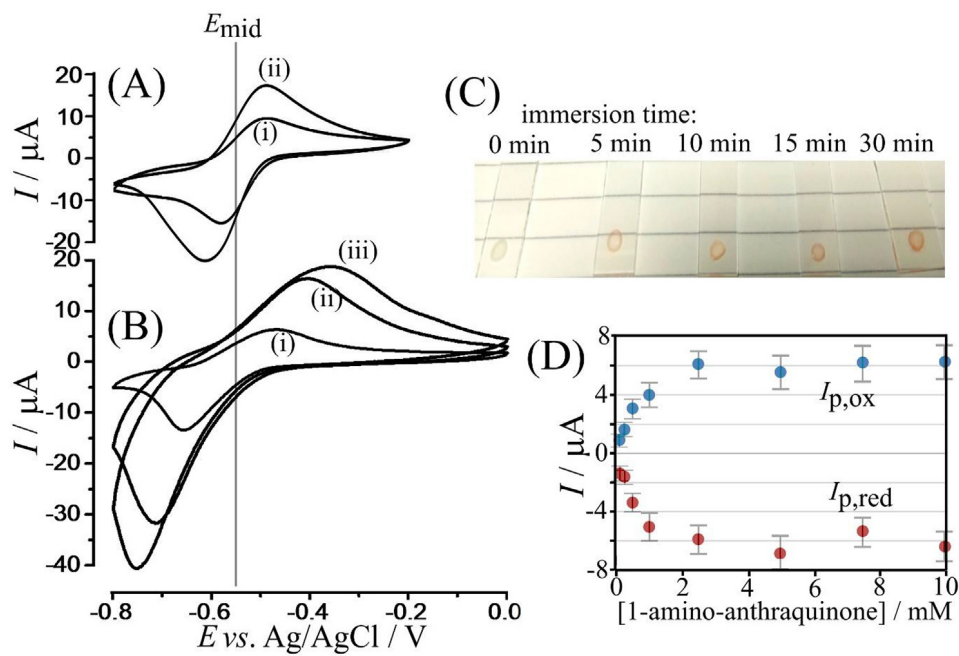

**Fig. 5.** (A) Cyclic voltammogram (scan rate 100 mVs$^{-1}$; in 0.1 M phosphate buffer pH 7) for 1-aminoanthraquinone extracted from 2.5 mM solution in cyclopentanone at (i) 1.9 μm thickness titanate film modified GCE and (ii) 3.3 μm thickness titanate film modified GCE. (B) Cyclic voltammogram (scan rate 100 mV s$^{-1}$; in (i) 0.1, (ii) 0.5, (iii) 1 M NaCl) for 1-aminoanthraquinone extracted from 2.5 mM solution in cyclopentanone at 1.9 μm thickness titanate film modified GCE. (C) Coloration as a function of time for the titanate film after immersion into the red-coloured 1-amino-anthraquinone. (D) Plot of anodic and cathodic peak currents *versus* concentration of 1-aminoanthraquinone in cyclopentanone solution (estimated error 10%).

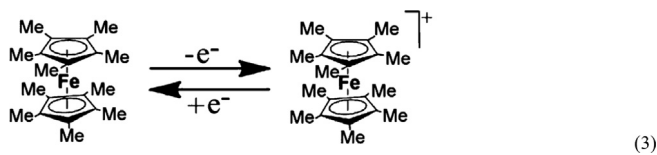

(3)

Fig. 6A shows typical data when the potential is cycled over nine consecutive potential cycles. A considerable decrease in the voltammetric response is observed, but the position of oxidation and reduction signal are consistent with the decamethylferrocene/decamethylferricenium redox system (see Fig. 6). The observed gradual loss of signal is likely to be associated with slow loss of the slightly water-soluble decamethylferricenium cation into the aqueous solution phase (or away from the electrochemical reaction zone). Data in Fig. 6B demonstrate the effect of titanate nanosheet host film thickness. Again, the increase in thickness appears to increase the peak currents for the voltammetric response. When systematically changing the concentration of the decamethylferrocene in cyclopentanone an increase in peak currents for both oxidation and reduction (first potential cycles, Fig. 6C and D) are observed. There is no clear plateau possibly due to cation expulsion during oxidation (not anion expulsion as in the case of anthraquinone reduction). In the Raman data (Fig. 6E) a rich set of vibrational peaks are seen for decamethylferrocene (ii) in addition to those for the titanate nanosheet host signals (i).

### 3.3. Binding of TetraphenylporphyrinatoMn(III)Cl into a titanate nanosheet host

Phorphyrin metal complexes are well known catalysts and electro-catalysts and therefore often immobilised into host materials [41,42]. Recently, tetraphenylporphyrin-Fe(III) complexes were immobilised into polymers of intrinsic microporosity to act as immobilised electrocatalysts at glassy carbon electrodes [43]. Here, the immobilisation and reactivity of tetraphenylporphyrin-Mn(III)Cl (or (TPPMnCl) into titanate nanosheet films is investigated. There have been several previous reports for TPPMnCl immobilisation and redox cycling in oil microdroplet deposits at electrode surfaces [44].

Data for TPPMnCl immobilised into titanate nanosheet films is summarised in Fig. 7. TPPMnCl is soluble in cyclopentanone giving a dark green coloration. When employing a 2.5 mM TPPMnCl solution and immersing a titanate nanosheet film, a colour change from white to green can be observed gradually over 30 min (see Fig. 7F). The voltammetric response for the immobilised TPPMnCl was first investigated immersed in aqueous 0.1 M phosphate buffer at pH 7. Fig. 7A shows consecutive voltammetric signals recorded at a scan rate of 100 mVs$^{-1}$. A reduction of TPPMnCl occurs followed by re-oxidation consistent with a one-electron process (Mn(III/II), see equation 4 with $L_1 = Cl^-$ and $L_2 = H_2O$). A decrease in voltammetric signal occurs most likely due to a chemical change such as the loss (or exchange) of the chloro-ligand upon reduction to TPPMn(II). Additional experiments were performed by intentionally adding NaCl into the aqueous phosphate buffer solution, but without

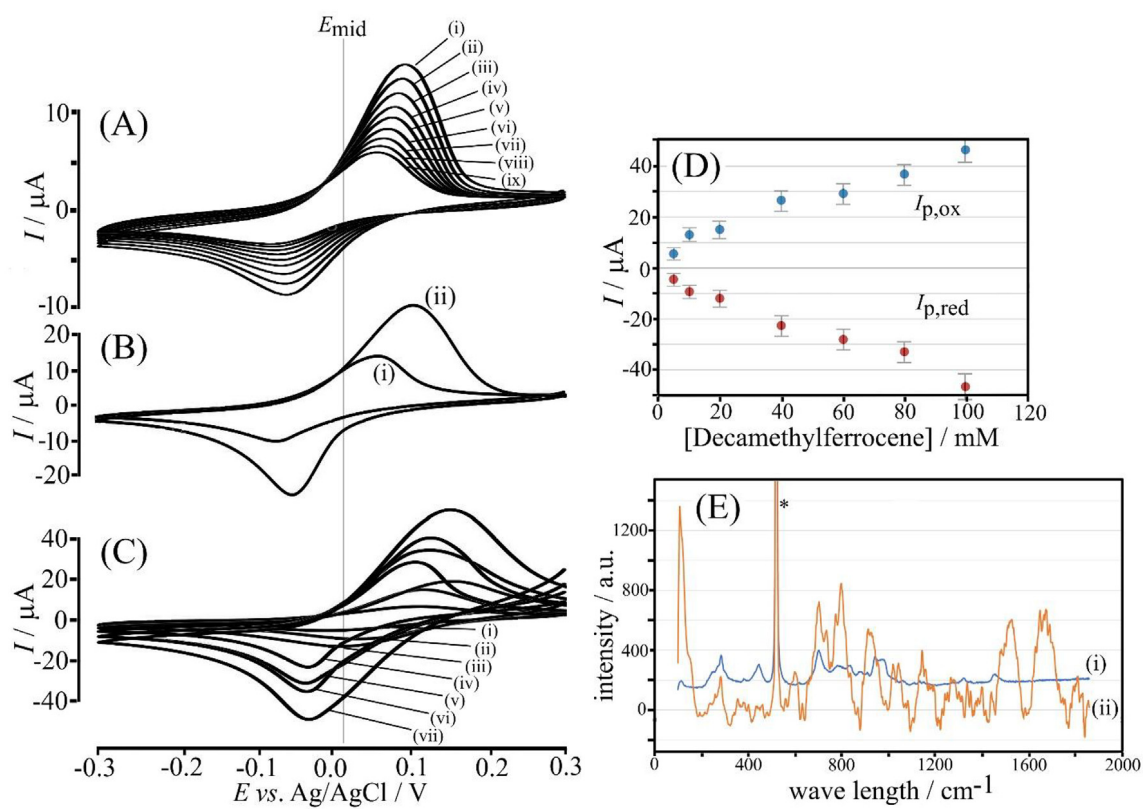

**Fig. 6.** (A) Cyclic voltammogram (nine continuous cycles; scan rate 100 mVs$^{-1}$; in 0.1 M phosphate buffer pH 7) for decamethylferrocene extracted from 40 mM solution in cyclopentanone into a 1.9 μm thickness titanate film modified GCE. (B) Cyclic voltammogram (scan rate 100 mVs$^{-1}$; in 0.1 M phosphate buffer pH 7) for decamethylferrocene extracted from 2.5 mM solution in cyclopentanone at (i) 1.9 μm thickness titanate film modified GCE and (ii) 3.3 μm thickness titanate film modified GCE. (C) Cyclic voltammogram (scan rate 100 mVs$^{-1}$; in 0.1 M phosphate buffer pH 7) for decamethylferrocene extracted from (i) 5, (ii) 10, (iii) 20, (iv) 40, (v) 60, (vi) 80, and (vii) 100 mM solution in cyclopentanone at 1.9 μm thickness titanate film modified GCE. (D) Plot of peak current versus concentration of decamethylferrocene in cyclopentanone solution (estimated error 10%). (E) Raman spectra of a pure titanate film (i blue) and a film with absorbed decamethylferrocene (ii, orange). The * symbols indicates a silicon substrate signal.

significant improvement in the signal stability. However, when performing the cyclic voltammetry experiment in the absence of phosphate and directly in aqueous 1 M NaCl (see Fig. 7B), the voltammetric response remains more stable. This can be attributed to the absence of phosphate as nucleophile and the presence of sufficiently high concentrations of chloride within the inter-lamellar space to prevent loss of chloride from the reduced TPPMn(II)Cl complex. In further experiments in aqueous 2 M NaCl a higher current response is seen and even an increase in the voltammetric response is observed with consecutive potential cycles (see Fig. 7C). The increase in current is reminiscent to increasing current responses in higher concentration of electrolyte, as seen for example for anthraquinone. In addition, there is a drift in the voltammetric responses with increasing potential cycle number to more negative potentials. This drift in potential could be indicative of a salt exchange process accompanying redox cycling over time, possibly linked to a change in the interfacial potential at the titanate nanosheet | aqueous electrolyte interface (*vide infra*). The improved

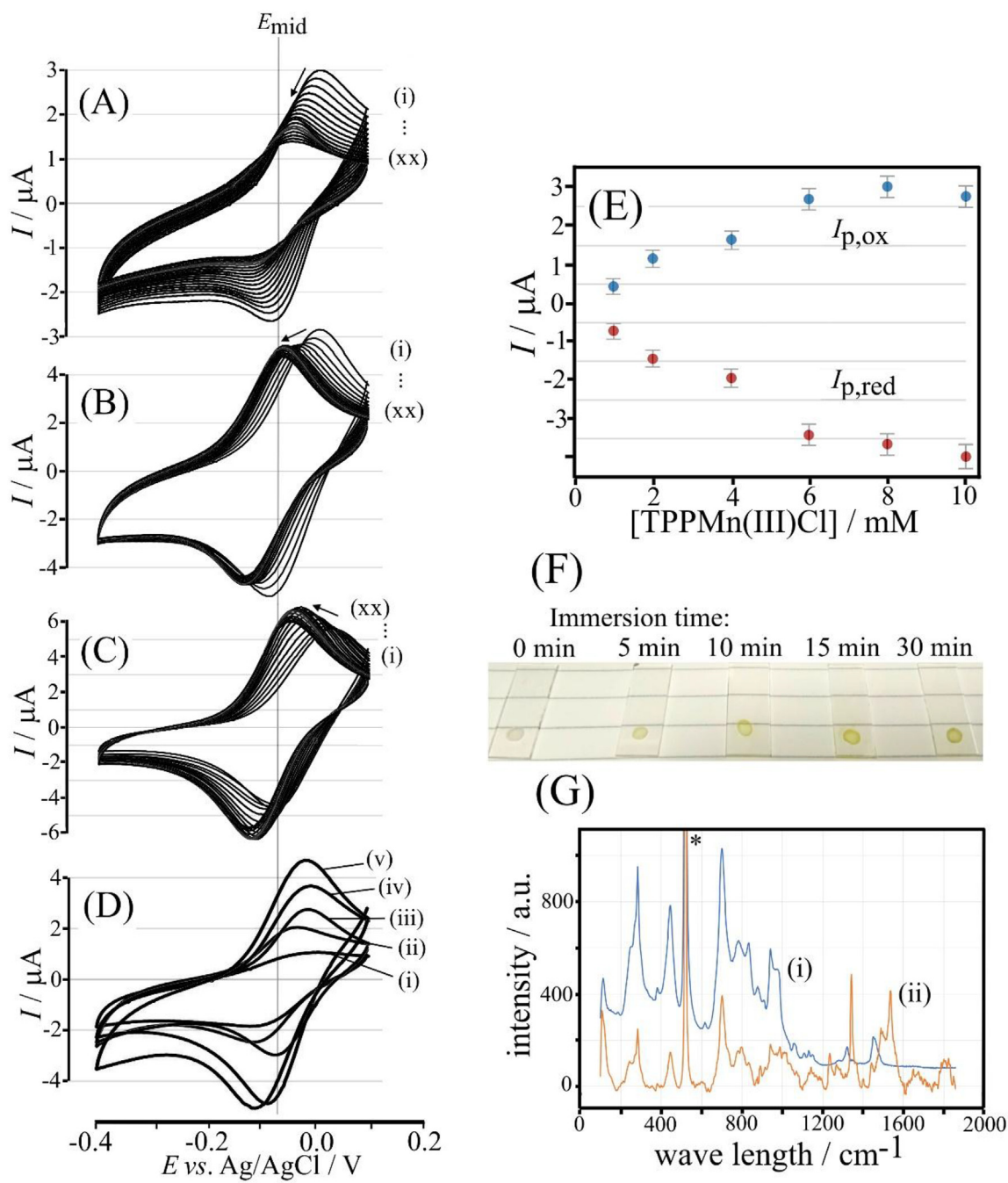

**Fig. 7.** (A) Cyclic voltammogram (twenty continuous cycles; scan rate 100 mV s$^{-1}$) for TPPMnCl extracted from 8 mM solution in cyclopentanone at 1.9 μm thickness titanate film modified GCE in 0.1 M phosphate buffer pH 7 as electrolyte. (B) As before in 1 M NaCl solution as electrolyte. (C) As before in 2 M NaCl solution as electrolyte. (D) Cyclic voltammogram (scan rate 100 mVs$^{-1}$; in 0.1 M phosphate buffer pH 7) for TPPMnCl extracted from (i) 1, (ii) 2, (iii) 4, (iv) 6, (v) 8 mM solution in cyclopentanone at 1.9 μm thickness titanate film modified GCE. (E) Plot of peak current *versus* concentration of TPPMnCl bound from cyclopentanone solution (estimated error 10%). (F) Coloration as a function of time of the titanate film after immersion into the 2.5 mM green-coloured TPPMnCl solution in cyclopentanone. (G) Raman spectra of a pure titanate film (i, blue) and film with absorbed TPPMnCl (ii, orange). The * symbols indicates a silicon substrate signal.

stability of the voltammetric response for TPPMnCl in the presence of chloride containing electrolyte can be linked to the presence of excess chloride in the titanate film to prevent chloro-ligand substitution (a process known for TPPMn(II)Cl in organic media [45]). When changing the concentration of TPPMnCl in cyclopentanone (see Fig. 7D and E) a systematic increase in voltammetric peak current is observed with an apparent plateau being reached may be at about 8 mM solution.

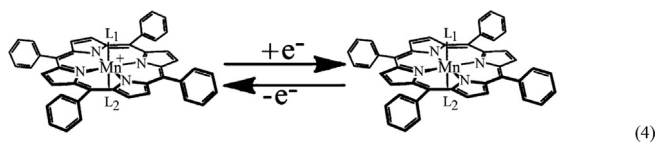

(4)

Raman data for pure titanate nanosheets and for TPPMnCl bound into titanate nanosheets are shown in Fig. 7G. Significant bands for embedded TPPMnCl are seen at approximately 1350 $cm^{-1}$ and around 1500 $cm^{-1}$. The latter group of peaks could be related to Raman peaks observed previously in organic solution [46]. The band at 1350 $cm^{-1}$ could be related to data reported for resonance Raman measurements in different solvents [47].

### 3.4. Binding of α-tocopherol derivatives into a titanate nanosheet host

An analytically relevant redox system is α-tocopherol (or vitamin E), which occurs naturally [48] and in domestic and pharmaceutical formulations [49]. α-Tocopherol is highly water-insoluble (see molecular structure in Fig. 8) and acts as anti-oxidant. Electrochemical properties have been investigated in organic solution [50,51] and for the oil phase surrounded by aqueous electrolyte [52] and for aqueous microemulsion systems [53]. The electrochemial oxidation of α-tocopherol in aqueous media at pH 7 occur at approximately 0.3 V vs. SCE [54]. This oxidation process is chemically highly irreversible. In addition to the process indicated in the reaction equation 5, also one-electron oxidation followed by radical dimerisation has been reported [54].

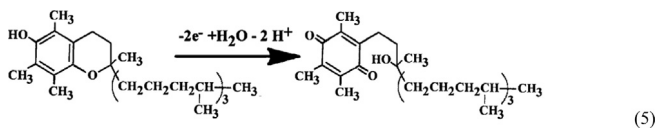

(5)

When immobilised into titanate nanosheet films from 2.5 mM α-tocopherol solution in cyclopentanone, the irreversible oxidation response is clearly observed (Fig. 8A). Doubling the titanate film thickness does lead to an approximate doubling of the anodic peak current in the voltammetric response. Only the first potential cycle is shown as the following potential cycles showed only insignificant peak currents (due to chemical irreversibility). Experiments were performed to investigate the effect of the organic solvent on the immobilisation process (see Fig. 8B). Each solvent clearly affects the immobilisation differently with ethanol, acetone, and dichloromethane giving higher current peaks. Chloroform appeared to be less suitable. Absorption α-tocopehol from cyclopentanone provided an intermediate current response.

Data in Fig. 8C show anodic peak current data (first potential cycle) for α-tocopherol oxidation after absorption from cyclopentanone. The increase in peak current is close to linear and occurs over a considerable range of concentrations. At the lower concentration limit, voltammetric peaks are too broad to be clearly detected under these conditions. Therefore practical analytical applications will require further optimisation and improvement of the current response. Raman spectroscopy data (Fig. 8D) show that the α-tocopherol immobilised into titanate nanosheet films only leads to minor changes in spectra with new broad bands at approximately 1500 $cm^{-1}$ and at 1800 $cm^{-1}$. The band at 1500 $cm^{-1}$ appears to be linked to α-tocopherol Raman spectra obtained in the pure compound [55].

### 3.5. Binding and reactivity of redox active species in a titanate nanosheet host

To summarise observations, it is possible to extract some general conclusions and mechanistic information from the data reported in this study. Highly hydrophobic redox active species are bound from organic solvents into titanate nanosheet host materials. This is likely to happen *via* a swelling process (compare [20]), in which the organic solvent enters the titanate inter-layer spacing and in this way allows slow ingress of the redox active guest. This explains the effect of the nature of the organic solvent on the process. Evaporation of the organic solvent leaves the redox system immobilised in the deposited titanate film.

The titanate nanosheet films compare in behaviour to hydrophobic oil microdroplet deposits containing redox active hydrophobic guests [56] with coupled electron transfer (at the titanate nanosheet | electrode interface) and ion transfer (at the titanate nanosheet | electrolyte solution interface). The apparent hydrophobic behaviour of titanate nanosheets is probably attributed to tetrabutylammonium cation derivatization of the surface of the nanosheets during the delapidation process. The electrochemical conversion of the hydrophobic guest species is likely to occur in a very thin layer close to the electrodes surface (see Fig. 9). Very slow diffusion of these molecules within a thin diffusion layer before and after conversion seems plausible. Bigger molecules such as TPPMnCl show markedly lower currents consistent with lower mobility. The diffusion layer is likely to be only a fraction of the thickness of the titanate host layer (see Fig. 9: "redox process layer").

The (generally observed) considerable effect of the titanate film thickness on the peak currents is indicative of another more mobile species being involved in the process. Ion transport through the titanate film and exchange with the outside solution is required for charge neutrality to be maintained. The amount of freely mobile ions in the titanate nanosheet layer will depend on the thickness, and therefore a thicker film might simply increase the amount of available freely mobile ions to sustain the voltammetric peak signal (see Fig. 9: "ion transport layer"). This also explains the increase in peak current with higher salt concentration in the outside electrolyte as associated with increased salt partitioning.

The improved chemical stability of TPPMnCl in high concentration NaCl seems very likely to be linked to suppression of chloride loss/ligand exchange after reduction (or suppression of exchange with other ligands such as phosphate). In contrast, decamethylferricenium does seem to escape from the redox process layer consistent with literature observations for solid state reactivity and dissolution of decamethylferricenium cations [40]. The absence of a significant pH effect on the voltammetric signal for the immobilised anthraquinone redox system could be linked to unbuffered conditions inside of the titanate nanosheet film (there is no internal buffer system to link to the outside aqueous buffer system). This is confirmed by the widening peak-to-peak separation for reduction and oxidation peaks for anthraquinone derivatives upon increasing the NaCl electrolyte concentration. During reduction of anthraquinone, the local pH inside of the nanosheet host increases and the reduction peak shifts negative. During oxidation the pH decreases and the oxidation peak shifts positive. The results probably also indicate that for anthraquinone reduction an exchange of anions occurs rather than the exchange of cations or protons (Fig. 9). Due to titanate nanosheet carrying multiple negative charges, these may also act as poly-anions with low mobility but high concentration. This could lead to competition between cation uptake at the interface titanate|aqueous electrolyte and swelling with partial poly-anion hydration. The titanate poly-anion concentration at the interface is always (intrinsically) high and therefore cation uptake could be slower. The potential drift of voltammetric responses (e.g. observed for TPPMnCl) may occur due to build up of interfacial charges during poly-anion swelling at the titanate|aqueous electrolyte interface (see illustration in Fig. 9). The interfacial potential between titanate film and outside aqueous solution phase may be poorly defined and drifting, as seen in the data for consecutively potential cycled TPPMnCl or decamethylferrocene in titanate film deposits. This also can be attributed to titanate nanosheet poly-anions swelling and partially transferring into the aqueous phase.

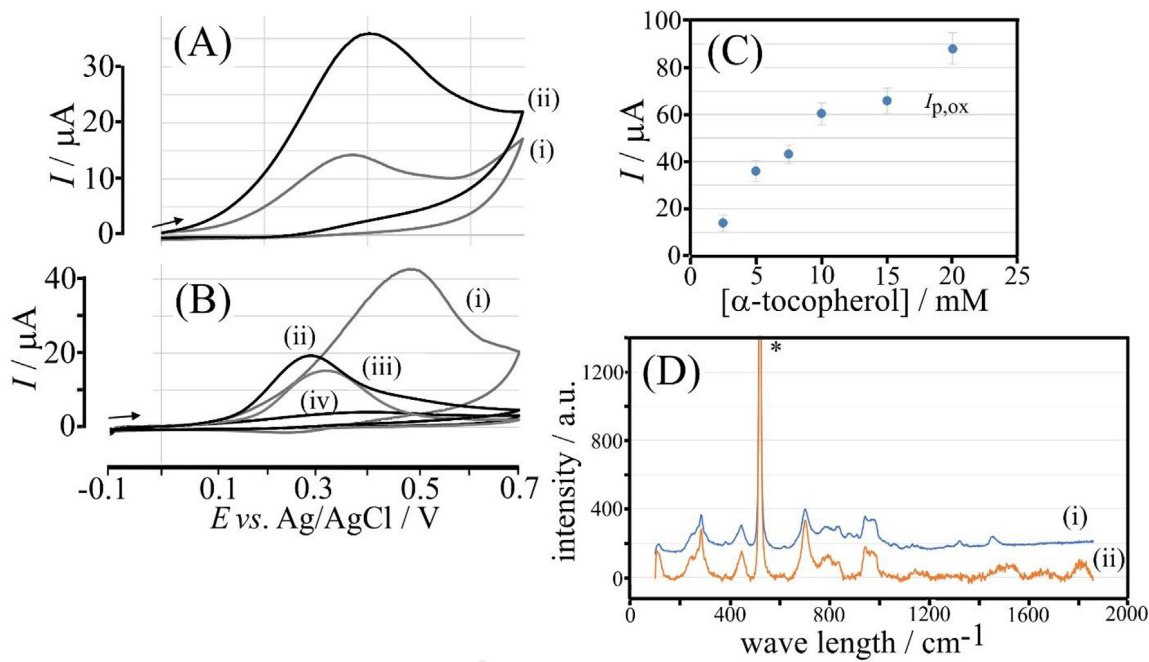

**Fig. 8.** (A) Cyclic voltammogram (scan rate 100 mVs$^{-1}$; in 0.1 M phosphate buffer pH 7) for α-tocopherol extracted from 2.5 mM solution in cyclopentanone at (i) 1.9 μm thickness titanate film modified GCE and (ii) 3.3 μm thickness titanate film modified GCE. (B) Cyclic voltammograms (scan rate 100 mVs$^{-1}$; in 0.1 M phosphate buffer pH 7) for α-tocopherol extracted from 2.5 mM solution in (i) dichloromethane, (ii) ethanol, (iii) acetone, (iv) chloroform at 1.9 μm thickness titanate film modified GCE. (C) Plot of peak current *versus* concentration of α-tocopherol in cyclopentanone solution (estimated error 10%). (D) Raman spectra of a pure titanate film (i, blue) and a film with absorbed α-tocopherol (ii, orange). The * symbols indicates a silicon substrate signal.

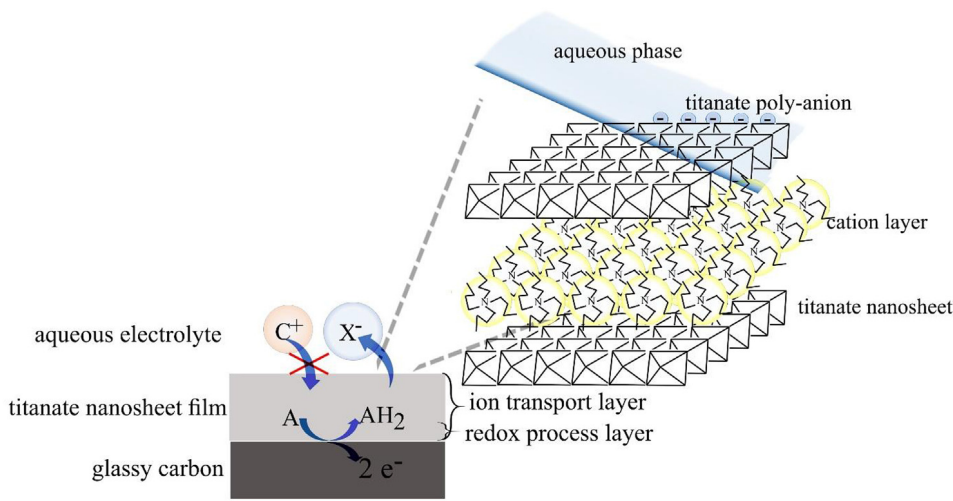

**Fig. 9.** Schematic depiction of a redox process A → AH$_2$ within a thin layer in a titanate nanosheet film associated with ion transport in the ion transport layer and with ion exchange at the titanate | aqueous electrolyte interface. The aqueous electrolyte is denoted C$^+$ and X$^-$. Uptake of C$^+$ is suggested to compete to exfoliation of titanate poly-anions.

The ability to readily bind hydrophobic molecules from organic solution directly into titanate nanosheet hosts seems novel and potentially important for analyte extraction followed by voltammetric detection. In this study, only relatively high concentrations of analyte (in the milli-molar range) were investigated and therefore for analytical applications further optimisation will be required to enhance voltammetric current signals (this could be achieved for example by increasing ionic strength, by more careful choice of the supporting electrolyte system, by design of the inter-lamellar space, and by pulse voltammetry). More generally, this study shows how 2D-nanosheet materials affect binding and reactivity of redox active analytes and there should be a wider search for more selective and more strongly binding 2D-nanosheet materials based on other related types of systems. The role of the alkylammonium cation used during exfoliation appears to be important. The interplay of swelling, absorption and binding, and transfer to an analytical detection device seem to be an attribute of lamellar or 2D materials and this could be further developed for routine analytical detection.

## 4. Conclusions

It has been demonstrated that highly lipophilic redox active species such as anthraquinone, 1-aminoanthraquinone, decamethylferrocene, TPPMnCl, and α-tocopherol are readily immobilised into titanate nanosheet film deposits at glassy carbon electrode surfaces. These redox species bind from organic solvents, in particular from cyclopentanone. The effect of the organic solvent on the swelling/absorption process suggests some involvement of the solvent by co-intercalation.

Electrochemical processes for titanate nanosheet immobilised species are similar to those reported for organic microphase environments and associated with electron transfer driven processes coupled to interfacial ion exchange. As a result, usually pH-sensitive processes, for example for anthraquinone reduction and for α-tocopherol oxidation, become pH-insensitive and associated with exchange of other types of electrolyte ions. Diffusion of redox active species such as anthraquinone and 1-aminoanthraquinone appears to be possible within the titanate nanosheets, but is suggested to be slow and contained within a thin zone close to the electrode surface. Diffusion of excess electrolyte within the inter-lamellar space appears to be linked to the process responsible for voltammetric characteristics. Redox processes for other types of species such as decamethylferrocene and TPPMnCl show gradual loss of peak currents, presumably due to either loss of redox active material away from the reaction zone or loss due to chemical instability (loss of chloride from the Mn(II) complex) based on the conditions within the titanate nanosheet film. α-Tocopherol is oxidised in a chemically irreversible manner, but shows a close to linear plot of anodic peak current versus concentration in a 1 mM–20 mM concentration range. Therefore probing of α-tocopherol concentration in organic oil media could be performed directly by binding into titanate nanosheet film electrodes. However, for analytical application the sensitivity needs to be improved substantially. The mechanism for electron transfer within the lamellar host also needs further study. Finally, the binding and mobility of guest species in the lamellar host deserves further investigation.

In the future, titanate nanosheet film deposits could be employed in separation and analytical detection based on extraction from organic solvent media. For this, for example the exfoliation cation could be modified to provide additional analyte selectivity and binding. Particularly useful could be the ability to respond to different types of solvents and to bind guests from a range of solvents ranging from highly polar to highly non-polar. Sensors based on a titanate film electrolyte may in future be able to operate in organic sample media based on natural oils, edible oils, industry oil samples, or in the gas phase. Colour evidence and Raman spectroscopy evidence for binding as well as electrochemical detection have been demonstrated. Therefore both electrochemical and non-electrochemical modes of detection in titanates may also be possible.

## Declaration of interest

There are no competing interests to declare by the authors.

## Acknowledgments

B.R.P. thanks to Indonesian Endowment (LPDP RI) for a PhD scholarship. W.T.W. acknowledges support from the Directorate of Collaboration and the International Program of Bogor Agricultural University for Short Term Research Program funding.

## References

[1] P. Wang, Z.B. Mai, Z. Dai, Y.X. Li, X.Y. Zou, Construction of Au nanoparticles on choline chloride modified glassy carbon electrode for sensitive detection of nitrite, Biosens. Bioelectron. 24 (2009) 3242–3324.
[2] S.J. Guo, S.J. Dong, Graphene nanosheet: synthesis, molecular engineering, thin film, hybrids, and energy and analytical applications, Chem. Soc. Rev. 40 (2011) 2644–2672.
[3] A. Sinha, Dhanjai, H.M. Zhao, Y.J. Huang, X.B. Lu, J.P. Chen, R. Jain, MXene: an emerging material for sensing and biosensing, Trac. Trends Anal. Chem. 105 (2018) 424–435.
[4] A. Sinha, Dhanjai, B. Tan, Y.J. Huang, H.M. Zhao, X.M. Dang, J.P. Chen, R. Jain, $MoS_2$ nanostructures for electrochemical sensing of multidisciplinary targets: a review, Trac. Trends Anal. Chem. 102 (2018) 75–90.
[5] M. Pumera, A.H. Loo, Layered transition-metal dichalcogenides ($MoS_2$ and $WS_2$) for sensing and biosensing, Trac. Trends Anal. Chem. 61 (2014) 49–53.
[6] B.Q. Yuan, C.Y. Xu, D.H. Deng, Y. Xing, L. Liu, H. Pang, D.J. Zhang, Graphene oxide/nickel oxide modified glassy carbon electrode for supercapacitor and nonenzymatic glucose sensor, Electrochim. Acta 88 (2013) 708–712.
[7] F.W. Li, D.R. MacFarlane, J. Zhang, Recent advances in the nanoengineering of electrocatalysts for $CO_2$ reduction, Nanoscale 10 (2018) 6235–6260.
[8] S. Bai, Y.J. Xiong, Recent advances in two-dimensional nanostructures for catalysis applications, Sci. Adv. Mater. 7 (2015) 2168–2181.
[9] G.H. Yang, C.Z. Zhu, D. Du, J.J. Zhu, Y.H. Lin, Graphene-like two-dimensional layered nanomaterials: applications in biosensors and nanomedicine, Nanoscale 7 (2015) 14217–14231.
[10] M.J. Wang, Z. Fan, L.Q. Yi, J.S. Xu, X.B. Zhang, Z.W. Tong, Construction of iron porphyrin/titanoniobate nanosheet sensors for the sensitive detection of nitrite, J. Mater. Sci. 53 (2018) 11403–11414.
[11] W.T. Wahyuni, B.R. Putra, C. Harito, D.V. Bavykin, F.C. Walsh, T.D. James, G. Kociok-Köhn, F. Marken, Electroanalysis in 2D-$TiO_2$ nanosheet hosts: electrolyte and selectivity effects in ferroceneboronic acid - saccharide binding, Electroanalysis 30 (2018) 1303–1310.
[12] D.V. Bavykin, F.C. Walsh, Titanate and Titania Nanotubes: Synthesis, Properties and Applications, Royal Society of Chemistry Nanoscience and Nanotechnology Monograph, London, 2009. ISBN: 978-1-84755-910-4.
[13] D.V. Bavykin, J.M. Friedrich, F.C. Walsh, Protonated titanates and $TiO_2$ nanostructured materials: synthesis, properties and applications, Adv. Mater. 18 (2006) 2807–2824.
[14] D.V. Bavykin, F.C. Walsh, The kinetics of alkali metal ion-exchange into nanotubular and nanofibrous titanates, J. Phys. Chem. C 111 (2007) 14644–14651.
[15] T. Sasaki, M. Watanabe, Semiconductor nanosheet crystallites of quasi-$TiO_2$ and their optical properties, J. Phys. Chem. B 101 (1997) 10159–10161.
[16] T. Sasaki, Y. Ebina, Y. Kitami, M. Watanabe, T. Oikawa, Two-dimensional diffraction of molecular nanosheet crystallites of titanium oxide, J. Phys. Chem. B 105 (2001) 6116–6121.
[17] N. Sakai, Y. Ebina, K. Takada, T. Sasaki, Electronic band structure of titania semiconductor nanosheets revealed by electrochemical and photoelectrochemical studies, J. Am. Chem. Soc. 126 (2004) 5851–5858.
[18] T. Sasaki, Molecular nanosheets of quasi-$TiO_2$: preparation and spontaneous reassembling, Supramol. Sci. 5 (1998) 367–371.
[19] T. Sasaki, M. Watanabe, H. Hashizume, H. Yamada, H. Nakazawa, Macromolecule-like aspects for a colloidal suspension of an exfoliated titanate. Pairwise association of nanosheets and dynamic reassembling process initiated from it, J. Am. Chem. Soc. 118 (1996) 8329–8335.
[20] C. Harito, D.V. Bavykin, M.E. Light, F.C. Walsh, Titanate nanotubes and nanosheets as a mechanical reinforcement of water-soluble polyamic acid: experimental and theoretical studies, Composites Part B 124 (2017) 54–63.
[21] A.S. Martins, C. Harito, D.V. Bavykin, F.C. Walsh, M.R.V. Lanza, Insertion of nanostructured titanates into the pores of an anodized $TiO_2$ nanotubes array by mechanically stimulated electrophoretic deposition, J. Mater. Chem. C 5 (2017) 3955–3961.
[22] H.B. Huang, Y.L. Ying, X.S. Peng, Graphene oxide nanosheet: an emerging star material for novel separation membranes, J. Mater. Chem. 2 (2014) 13772–13782.
[23] F. Marken, J.D. Watkins, A.M. Collins, Ion-transfer- and photo-electrochemistry at liquid | liquid | solid electrode triple phase boundary junctions: perspectives, Phys. Chem. Chem. Phys. 13 (2011) 10036–10047.
[24] T. Sasaki, M. Watanabe, Osmotic swelling to exfoliation. Exceptionally high degrees of hydration of a layered titanate, J. Am. Chem. Soc. 120 (1998) 4682–4689.
[25] L.Z. Wang, T. Sasaki, Titanium oxide nanosheets: graphene analogues with versatile functionalities, Chem. Rev. 114 (2014) 9455–9486.
[26] G. Matricali, M.M. Dieng, J.F. Dufeu, M. Guillou, Electrochemistry in solid-state

[26] of 9,10-anthraquinone-anthraquinol couple — use in negative electrode-reactions of secondary electrochemical generators, Electrochim. Acta 21 (1976) 943–952.
[27] A. Ullah, A. Rauf, U.A. Rana, R. Qureshi, M.N. Ashiq, H. Hussain, H.B. Kraatz, A. Badshah, A. Shah, pH dependent electrochemistry of anthracenediones at a glassy carbon electrode, J. Electrochem. Soc. 162 (2015) H157–H163.
[28] L. Devlin, M. Jamal, K.M. Razeeb, Novel pH sensor based on anthraquinone-ferrocene modified free standing gold nanowire array electrode, Anal. Methods 5 (2013) 880–884.
[29] F. Scholz, Electroanalytical Methods, Springer, Berlin, 2010, p. 69.
[30] M.A. Ghanem, I. Kocak, A. Al-Mayouf, M. AlHoshan, P.N. Bartlett, Covalent modification of carbon nanotubes with anthraquinone by electrochemical grafting and solid phase synthesis, Electrochim. Acta 68 (2012) 74–80.
[31] J.D. Watkins, K. Lawrence, J.E. Taylor, T.D. James, S.D. Bull, F. Marken, Carbon nanoparticle surface electrochemistry: high-density covalent immobilisation and pore-reactivity of 9,10-anthraquinone, Electroanalysis 23 (2011) 1320–1324.
[32] G.G. Wildgoose, M. Pandurangappa, N.S. Lawrence, L. Jiang, T.G.J. Jones, R.G. Compton, Anthraquinone-derivatised carbon powder: reagentless voltammetric pH electrodes, Talanta 60 (2003) 887–893.
[33] E.V. Milsom, J. Novak, S.J. Green, X.H. Zhang, S.J. Stott, R.J. Mortimer, K. Edler, F. Marken, Layer-by-layer deposition of open-pore mesoporous $TiO_2$-Nafion (R) film electrodes, J. Solid State Electrochem. 11 (2007) 1109–1117.
[34] T. Gao, H. Fjellvag, P. Norby, Raman scattering properties of a protonic titanate $H_xTi_{2-x/4}\square_{0-x/4} H_2O$ (x = 0.7) with lepidocrocite-type layered structure, J. Phys. Chem. B 112 (2008) 9400–9405.
[35] T. Gao, H. Fjellvag, P. Norby, Protonic titanate derived from $Cs_xTi_{2-x/2}Mg_{x/2}O_4$ (x = 0.7) with lepidocrocite- type layered structure, J. Mater. Chem. 19 (2009) 787–794.
[36] K.K. Lehmann, J. Smolarek, O.S. Khalil, L. Goodman, Vibrational assignments for the Raman and phosphorescence spectra of 9,10-anthraquinone and 9,10-anthraquinone-$d_8$, J. Phys. Chem. 83 (1979) 1200–1205.
[37] F. Stenman, Raman scattering from powdered 9,10-anthraquinone, J. Chem. Phys. 51 (1969) 3413–3415.
[38] K. Peckova, H. FiizAyyidjz, M. Topkafa, H. Kara, M. Ersoz, J. Barek, Polarographic and voltammetric determination of trace amounts of 2-aminoanthraquinone, Chem. Anal. 52 (2007) 989–1001.
[39] V. Kratochvil, M. Nepras, Anthraquinone dystuffs.14. Protonation of 9,10-anthraquinone derivatives, Collect. Czech Chem. Commun. 37 (1972) 1533–1535.
[40] A.M. Bond, F. Marken, Mechanistic aspects of the electro and ion-transport processes across the electrode solid solvent (electrolytes) interface of microcrystalline decamethylferrocene attached mechanically to a graphite electrode, J. Electroanal. Chem. 372 (1994) 125–135.
[41] G.F. Manbeck, E. Fujita, A review of iron and cobalt porphyrins, phthalocyanines and related complexes for electrochemical and photochemical reduction of carbon dioxide, J. Porphyr. Phthalocyanines 19 (2015) 45–64.
[42] C. Costentin, M. Robert, J.M. Saveant, Molecular catalysis of electrochemical reactions, Curr. Opinion Electrochem. 2 (2017) 26–31.
[43] Y.Y. Rong, R. Malpass-Evans, M. Carta, N.B. McKeown, G.A. Attard, F. Marken, High density heterogenisation of molecular electrocatalysts in a rigid intrinsically microporous polymer, Electrochem. Commun. 46 (2014) 26–29.
[44] M.J. Bonne, C. Reynolds, S. Yates, G. Shul, J. Niedziolka, M. Opallo, F. Marken, The electrochemical ion-transfer reactivity of porphyrinato metal complexes in 4-(3-phenylpropyl)pyridine vertical bar water systems, New J. Chem. 30 (2006) 327–334.
[45] G.J. Foran, R.S. Armstrong, M.J. Crossley, P.A. Lay, Effect of electrolytes concentration on axial anion ligation in manganese(III) meso-tetraphenylporphyrin chlorides, Inorg. Chem. 31 (1992) 1463–1470.
[46] R.R. Gaughan, D.F. Shriver, L.J. Boucher, Resonance Raman-spectroscopy of manganese(III) tetraphenylporphyrin halides, Proc. Natl. Acad. Sci. Unit. States Am. 72 (1975) 433–436.
[47] S.C. Jeoung, D. Kim, D.W. Cho, Transient resonance Raman spectroscopic studies of some paramagnetic metalloporphyrins: effects of axial ligand on charge-transfer and photoreduction processes, J. Raman Spectrosc. 31 (2000) 319–330.
[48] S.N. Robledo, V.G.L. Zachetti, M.A. Zon, H. Fernandez, Quantitative determination of tocopherols in edible vegetable oils using electrochemical ultramicrosensors combined with chemometric tools, Talanta 116 (2013) 964–971.
[49] M. Sys, B. Svecova, I. Svancara, R. Metelka, Determination of vitamin E in margarines and edible oils using square wave anodic stripping voltammetry with a glassy carbon paste electrode, Food Chem. 229 (2017) 621–627.
[50] T. Kondo, K. Sakai, T. Watanabe, Y. Einaga, M. Yuasa, Electrochemical detection of lipophilic antioxidants with high sensitivity at boron-doped diamond electrode, Electrochim. Acta 95 (2013) 205–211.
[51] M. Coataena, A. Darchen, D. Hauchard, Electroanalysis at ultramicroelectrodes of oils and fats - application to the determination of Vitamin E, Sens. Actuators B-Chem. 76 (2001) 539–544.
[52] W.W. Yao, H.M. Peng, R.D. Webster, Electrochemistry of α-tocopherol (vitamin E) and α-tocopherol quinone films deposited on electrode surfaces in the presence and absence of lipid multilayers, J. Phys. Chem. C 113 (2009) 21805–21814.
[53] M.A. Ghanem, R.G. Compton, B.A. Coles, A. Canals, F. Marken, Microwave enhanced electroanalysis of formulations: processes in micellar media at glassy carbon and at platinum electrodes, Analyst 130 (2005) 1425–1431.
[54] A.J. Wain, J.D. Wadhawan, R.R. France, R.G. Compton, Biphasic redox chemistry of α-tocopherol: evidence for electrochemically induced hydrolysis and dimerization on the surface of and within femtolitre droplets immobilised onto graphite electrodes, Phys. Chem. Chem. Phys. 6 (2004) 836–842.
[55] J.R. Beattie, C. Maguire, S. Gilchrist, L.J. Barrett, C.E. Cross, F. Possmayer, M. Ennis, J.S. Elborn, W.J. Curry, J.J. McGarvey, B.C. Schock, The use of Raman microscopy to determine and localize vitamin E in biological samples, Faseb. J. 21 (2007) 766–776.
[56] C.E. Banks, T.J. Davies, R.G. Evans, G. Hignett, A.J. Wain, N.S. Lawrence, J.D. Wadhawan, F. Marken, R.G. Compton, Microdroplet review Electrochemistry of immobilised redox droplets: concepts and applications, Phys. Chem. Chem. Phys. 5 (2003) 4053–4069.